\documentclass[preprint, superscriptaddress]{revtex4-2}

\usepackage{amssymb,graphicx,color,amsmath,xfrac,bm,stmaryrd,trimclip} 
\usepackage[utf8x]{inputenc}
\definecolor{lightblue}{rgb}{0.17,0.39,1}
\usepackage[bookmarks, colorlinks=true, breaklinks]{hyperref}
\hypersetup{linkcolor=lightblue, citecolor=lightblue, filecolor=black, urlcolor=lightblue}

\begin{document} 

\title{{High-field-stabilized reentrant superconductivity in infinite-layer nickelate thin films}}




\author{Km Rubi$^\dagger$}
\email[email:] {rubi@lanl.gov}
\affiliation{National High Magnetic Field Laboratory, Los Alamos National Laboratory, Los Alamos, NM 87545, USA}

\author{King Yau Yip}
\thanks{These authors contributed equally.}
\affiliation{Department of Physics, Faculty of Science, National University of Singapore, Singapore 117551, Republic of Singapore}

\author{Elizabeth Krenkel}
\affiliation{Center for Integrated Nanotechnologies, Los Alamos National Laboratory, Los Alamos, NM 87545, USA}

\author{Nurul Fitriyah}
\affiliation{Department of Physics, Faculty of Science, National University of Singapore, Singapore 117551, Republic of Singapore}
\affiliation{Department of Physics, Faculty of Science and Technology, Universitas Airlangga, Surabaya 60115, East Java, Indonesia}

\author{Xing Gao}
\affiliation{Department of Physics, Faculty of Science, National University of Singapore, Singapore 117551, Republic of Singapore}

\author{Saurav Prakash}
\affiliation{Department of Physics, Faculty of Science, National University of Singapore, Singapore 117551, Republic of Singapore}

\author{S. Lin Er Chow}
\affiliation{Department of Physics, Faculty of Science, National University of Singapore, Singapore 117551, Republic of Singapore}

\author{Chi Sin Tang}
\affiliation{Singapore Synchrotron Light Source (SSLS), National University of Singapore, 5 Research Link, Singapore 117603, Republic of Singapore}

\author{Tsz Fung Poon}
\affiliation{Department of Physics, The Chinese University of Hong Kong, Shatin, Hong Kong, China}

\author{Swee K. Goh}
\affiliation{Department of Physics, The Chinese University of Hong Kong, Shatin, Hong Kong, China}

\author{Sean M. Thomas}
\affiliation{MPA-Q, Los Alamos National Laboratory, Los Alamos, NM 87545, USA}

\author{Adam P. Dioguardi}
\affiliation{MPA-Q, Los Alamos National Laboratory, Los Alamos, NM 87545, USA}

\author{Oscar E. Ayala-Valenzuela}
\affiliation{National High Magnetic Field Laboratory, Los Alamos National Laboratory, Los Alamos, NM 87545, USA}

\author{Mark B. H. Breese}
\affiliation{Singapore Synchrotron Light Source (SSLS), National University of Singapore, 5 Research Link, Singapore 117603, Republic of Singapore}
\affiliation{Department of Physics, Faculty of Science, National University of Singapore, Singapore 117551, Republic of Singapore}

\author{Mun K. Chan}
\affiliation{National High Magnetic Field Laboratory, Los Alamos National Laboratory, Los Alamos, NM 87545, USA}

\author{David Graf}
\affiliation{National High Magnetic Field Laboratory, Florida State University, Tallahassee, Florida 32310, USA}

\author{A. Ariando}
\email [email:] {ariando@nus.edu.sg}
\affiliation{Department of Physics, Faculty of Science, National University of Singapore, Singapore 117551, Republic of Singapore}

\author{Neil Harrison}
\affiliation{National High Magnetic Field Laboratory, Los Alamos National Laboratory, Los Alamos, NM 87545, USA}

\begin{abstract} 

\textbf {Magnetic fields typically suppress superconductivity through Pauli and orbital limiting effects. However, there are rare instances of magnetic-field–induced superconductivity, as observed in Chevrel-phase compounds~\cite{meul1984}, organic conductors~\cite{uji2001,balicas2001}, uranium-based heavy-fermion systems~\cite{levy2005,ran2019,helm2024,lewin2025}, and moiré graphene~\cite{cao2021}---although these materials possess inherently low superconducting transition temperatures ($T_{\rm c}$). Here, we demonstrate high-field–stabilized superconductivity in a class of materials recently shown to have significantly higher $T_{\rm c}$ values (up to 40~K): the infinite-layer nickelates~\cite{chow2025}. We show that both the low-field and high-field superconducting states can be understood in terms of a field-compensation mechanism, better known as the Jaccarino–Peter effect~\cite{jaccarino1962}. These findings demonstrate the possibility of achieving substantially enhanced upper critical fields in high-temperature superconductors.}
\end{abstract} 
\date{\today}	
\maketitle	

There is considerable technological interest in superconducting state that remains robust in high magnetic fields ~\cite{uglietti2019,mitchell2021,shiltsev2021}. One significant challenge is that strong magnetic fields are typically detrimental to superconductivity. This occurs via two primary mechanisms: Pauli pair breaking~\cite{clogston1962,chandrasekhar1962}, where the Zeeman interaction aligns electron spins and disrupts spin-singlet Cooper pairs, and orbital pair breaking, where magnetic flux penetrates the superconductor in the form of vortices, ultimately destroying phase coherence when the vortex density becomes too high~\cite{tinkham1996}.

One possible way to circumvent the detrimental effects of magnetic fields on superconductivity is to exploit situations in which the interplay between conduction electrons and magnetism gives rise to magnetic-field–induced superconducting states~\cite{meul1984,uji2001,balicas2001,levy2005,ran2019np,cao2021,helm2024,lewin2025}. Discovered in Chevrel-phase compounds~\cite{meul1984}, field-induced superconductivity has so far been restricted to materials with intrinsically low superconducting transition temperatures ($T_{\rm c}$). This limitation is often attributed to factors such as conventional electron–phonon-mediated pairing~\cite{meul1984} or the presence of narrow electronic bandwidths~\cite{levy2005,ran2019,cao2021}, which enhance electronic correlations but typically suppress $T_{\rm c}$.

Here we demonstrate that infinite-layer nickelates~\cite{li2019,sun2023,zhou2025,chow2025} provide an example of a magnetic-field--reentrant superconducting state persisting to very strong magnetic fields occurring in a material that also hosts high-temperature superconductivity comparable to that of the cuprates~\cite{keimer2015}. Specifically, we observe this phenomenon in 4-7~nm--thick films of (Sm$_{1-x-y-z}$Eu$_x$Ca$_y$Sr$_z$)NiO$_2$, hereafter referred to as SECNO, which were recently shown to exhibit transition temperatures approaching $T_{\mathrm{c}} \!\approx\! 40$~K~\cite{chow2025}---exceeding that of the first cuprate system in which high-$T_{\mathrm{c}}$ superconductivity was originally discovered~\cite{bednorz1986}. 

\begin{figure*}[!t] 
\begin{center}
\includegraphics[width=1.0\linewidth]{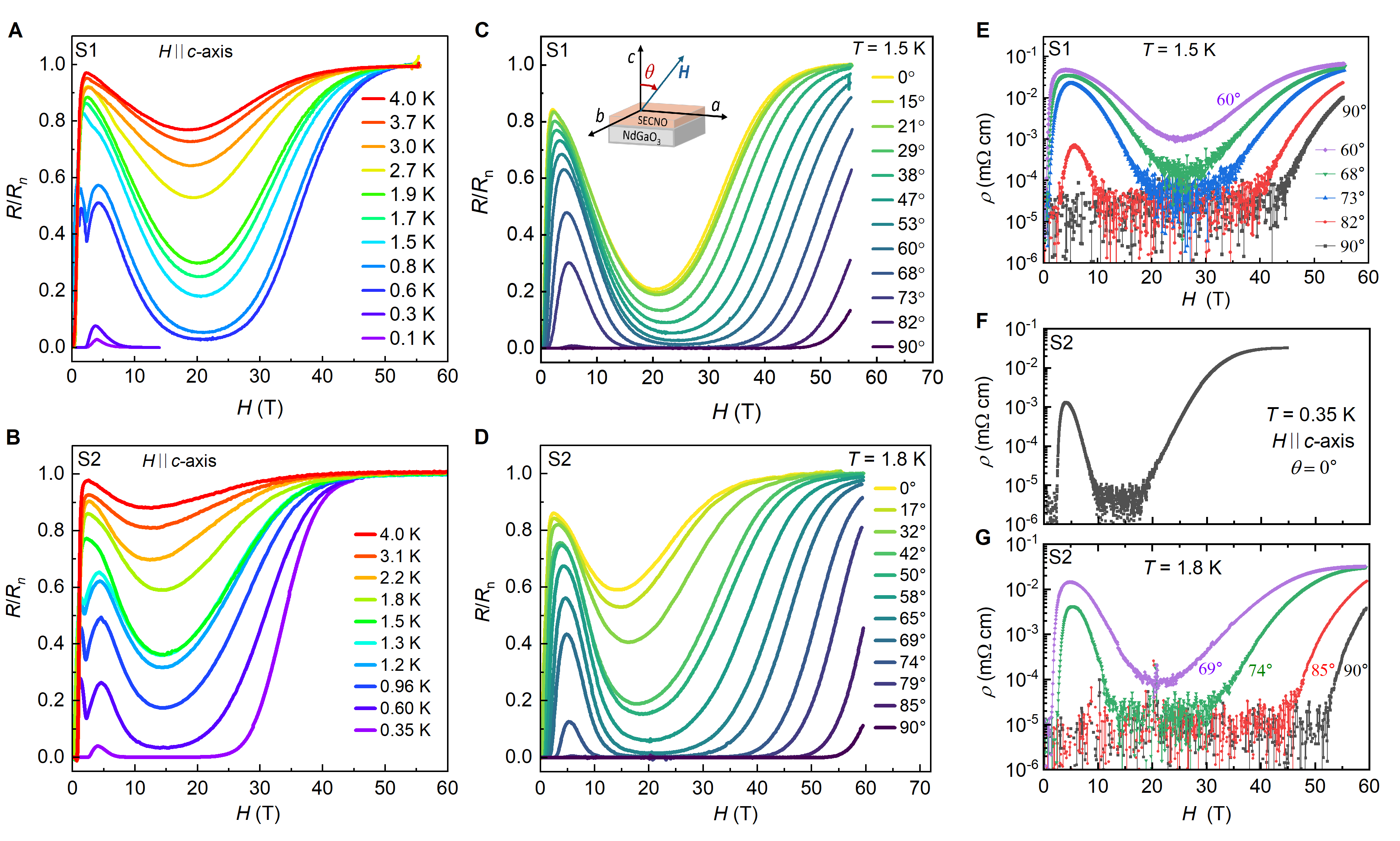}
\textsf{\caption{
{
\textbf{Re-entrant superconductivity at high magnetic fields in SECNO superconductors} (A) Measured renormalized electrical resistivity $R/R_{\rm n}$ versus magnetic field $H$ of sample S1 ($T_{\rm c}=$~9.6~K) at different temperatures $H$ applied along the $c$ axis. (B) Similar data for samples S2 ($T_{\rm c}=$~11.7~K). (C) and (D), $R/R_{\rm n}$ measured in both samples at constant temperature for different angles $\theta$ between the applied field and the $c$-axis. {The inset in C is a schematic of \(H\) alignment, moving away from the out-of-plane \(c\)-axis (\(\theta = 0^\circ \)) toward the in-plane \(a\)-axis (\(\theta = 90^\circ \))}. (E - G) Absolute resistivity $\rho$ plotted on a logarithmic scale, reveal it to drop by roughly four orders of magnitude. For compositions, see Table 1 in the Methods. 
}
}
\label{resistivity}}
\end{center}
\vspace{-0.2cm}
\end{figure*}

Figure~\ref{resistivity} shows the experimental evidence for reentrant superconductivity in resistance curves measured for two SECNO samples, S1 and S2, with relatively low transition temperatures of \(T_{\rm c} \approx 9.6~\mathrm{K}\) and 11.7~K, respectively (see Methods and Extended Table \ref{table1} for exact composition and film thickness). The resistance is measured as a function of magnetic field \(H\) at various temperatures \(T\), with \(H\) applied along the crystalline \(c\)-axis (Figs.~\ref{resistivity}A and B). Salient features of both samples include a sharp transition from the superconducting to the resistive state at \(H \approx 1~\mathrm{T}\), followed by a small dip in resistivity near \(H \approx 2~\mathrm{T}\) for \(T < 2~\mathrm{K}\). This dip coincides with the spin-flop transition of the antiferromagnetically ordered Nd moments in the NdGaO$_3$ substrate into a polarized state (see Extended Data Fig.~\ref{magcurves} and Methods). At higher fields, a broad resistivity minimum emerges—centered around \(H \approx 20~\mathrm{T}\) and \(15~\mathrm{T}\) for samples S1 and S2, respectively—which shifts systematically to higher fields as the tilt angle $\theta$ between \(H\) and the crystalline \(c\)-axis increases (see Figs.~\ref{resistivity}C and D, and Extended Data Fig.~\ref{lowerTangular}). At sufficiently large fields, the resistivity rises and saturates at a nearly constant value, consistent with the normal state.

Suggestive of a reentrant superconducting state involving a substantial fraction of the thin nickelate films, the resistivity plotted on a logarithmic scale drops by several orders of magnitude (see Figs.~\ref{resistivity}E--G) and becomes dominated by noise at sufficiently low temperatures, \( T \lesssim 0.35~\mathrm{K} \) at \( \theta = 0^\circ \),  at angles above \( \theta \sim 70^\circ \) at \( T = 1.5~\mathrm{K} \) in sample S1, or at angles above \( \theta \sim 74^\circ \) (\( 26^\circ \)) at \( T = 1.8~\mathrm{K} \) (\( 0.6~\mathrm{K} \)) in sample S2. The magnitude of the resistance drop at high fields is comparable to that observed in the low-field superconducting state at \( H < 1~\mathrm{T} \). Consistent with the presence of superconducting screening currents in the vanishing-resistivity regime of the thin films, radio-frequency (\(\approx\!20~\mathrm{MHz}\)) inductive measurements on sample S1 -- performed using a proximity-detector tank circuit~\cite{mikitik2025} adapted for pulsed magnetic fields~\cite{altarawneh2009} at \( T = 0.5~\mathrm{K} \) (see Extended Data Fig.~\ref{pdo}) -- reveal that the vanishing-resistivity state in high magnetic fields is accompanied by a diamagnetic shift associated with screening currents. The magnitude of this shift at high angles is comparable to that of the low-magnetic field ($H<$~1~T) superconducting state, providing compelling evidence of bulk superconductivity in high-field regimes.



\begin{figure*}[!t] 
\begin{center}
\includegraphics[width=0.9\linewidth]{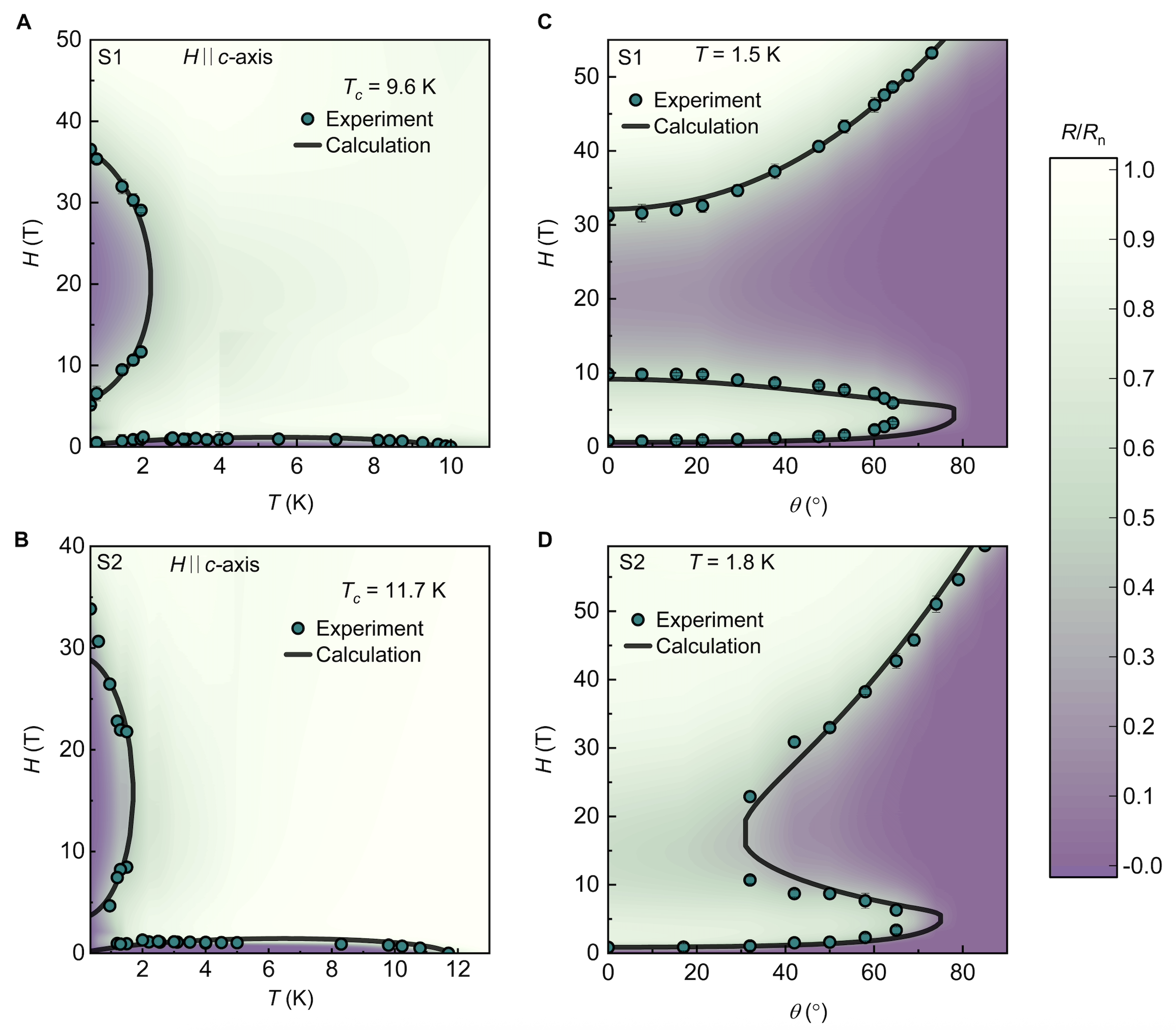}
\textsf{\caption{
{\textbf{High-field reentrant superconductivity in SECNO: Phase diagrams and evolution of upper critical field with temperature and tilt angle.} Phase diagrams of critical magnetic field versus temperature of (A) S1 ($T_{\rm c}=$~9.6~K) and (B) S2 ($T_{\rm c}=$~11.7~K) for $H$ parallel to the $c$-axis. Phase diagrams of critical magnetic field versus angle $\theta$ of (C) S1 at $T=$~1.5~K and (D) S2 at $T=$~1.8~K. Symbols are experimentally determined phase boundaries delineated as the midpoints of the resistive transitions and lines are calculated phase boundaries using a modified WHH model. 
}}
\label{phasediagram}}
\end{center}
\vspace{-0.2cm}
\end{figure*}

Colored contour plots in Fig.~\ref{phasediagram} indicate regions where the resistivity vanishes (purple) and where it reaches its normal-state value (light green). Following the convention introduced in the first discovery of magnetic-field–reentrant superconductivity in Chevrel-phase compounds~\cite{meul1984,rossel1985}, \(T_{\rm c}\) and \(H_{\rm c}\) data points refer to the midpoints of the resistive transitions, where the resistivity is halfway between zero and its saturated normal-state value. The \(T\)--\(H\) phase diagrams (Figs.~\ref{phasediagram}A and~B) clearly reveal two distinct superconducting phases: a low-field superconducting state and  a high-field–induced reentrant superconducting state. These two phases merge as the direction of \(H\) is rotated toward the in-plane \(a\)-axis (Figs.~\ref{phasediagram}C and~D, and Extended Data Fig.~\ref{lowerTangular}B), with superconductivity persisting at least up to 60~T (50~T) at \(T = 0.6~\mathrm{K}\) (1.5 and 1.8~K).

The mechanisms underlying the reentrant superconductivity have been the subject of extensive debate. Proposed explanations include compensation of the applied field by an internal exchange field~\cite{meul1984, balicas2001, helm2024}, electron pairing mediated by spin fluctuations~\cite{aoki2001, levy2005, Huy2007}, field-enhanced spin reorientation or spin-triplet pairing~\cite{levy2005, ran2019}, field-induced magnetic transitions and Fermi-surface reconstructions~\cite{Wu2025}, and Landau-level broadening near the quantum limit~\cite{Rasolt1992}.  Given the presence of rare-earth elements Sm and Eu within the nickelate films, or Nd within the substrate, it is plausible that field-induced reentrant superconductivity in SECNO arises from significant  {exchange interactions} $J_{\rm cf}$ between the conduction electrons and localized rare-earth magnetic moments~\cite{jaccarino1962}. In magnetic ion-containing superconductors~(e.g., Eu-containing Chevrel-phase compounds~\cite{fischer1975, meul1984} and Fe-based organic conductors~\cite{uji2001, balicas2001}), the conduction electrons experience a negative internal exchange field $H_J\sim J_{\rm cf}/\mu_{\rm B}$ from the magnetic ions (where $\mu_{\rm B}$ is the Bohr magneton). Because the internal exchange field aligns opposite to the applied magnetic field, the two can partially or completely cancel each other, thereby reducing the Pauli pair-breaking effect of the external field and allowing a magnetic-field–induced superconducting phase to appear, centered around \(H = |H_J|\). This magnetic-field compensation mechanism was predicted long before reentrant superconductivity was discovered and became known as the Jaccarino–Peter effect~\cite{jaccarino1962}.

Clues to the relevant mechanism for reentrant superconductivity in SECNO are provided by striking similarities to the Chevrel-phase compounds in which reentrant superconductivity was  first discovered~\cite{meul1984,rossel1985} when $H\|c$. These similarities are evident in both the resistivity curves (see Fig.~\ref{resistivity}) and the corresponding phase diagrams (see Fig.~\ref{phasediagram}), constructed using the same midpoint-of-transition criterion. In the Chevrel-phase compounds, the combined effects of the orbital critical field \(H_{\rm c}^{\rm orb}\) and spin–orbit scattering (parameterized by \(\lambda_{\rm SO}\)) are known to shift the reentrant superconducting state down in magnetic field so that its optimal (field-stabilized) $T_{\rm c}$ occurs at a magnetic field lower than \(|H_J|\)~\cite{fischer1975}. These effects are treated using the Werthamer–Helfand–Hohenberg (WHH) model~\cite{werthamer1966}, which is a microscopic model describing the upper critical field of conventional type-II superconductors in the dirty limit. Incorporating \(|H_J|\) into the WHH model reproduces several key features of the magnetic field–temperature phase diagram in Chevrel-phase compounds~\cite{rossel1985}. First, the unusually low critical magnetic field \(H_{\rm c} \sim 1~\mathrm{T}\) of the low-field superconducting phase~\cite{rossel1985,meul1984}, which can be understood as a consequence of the rapid polarization of the localized magnetic moments (Eu$^{2+}$ for which the angular momentum quantum number is $J = \tfrac{7}{2}$, and the Landé \(g\)-factor is $g_J = 2$) whose internal exchange fields destabilize the low-field state. Second, the dome-shaped reentrant superconducting phase that emerges once the  magnetic moments are fully polarized by the external magnetic field. The reduced transition temperature of the reentrant phase relative to that at zero field reflects a delicate interplay among competing energy scales in strong magnetic fields.

We find that the WHH model~\cite{fischer1975,rossel1985}, modified to include \(|H_J|\)~\cite{fischer1975,rossel1985}, provides an equally good description of the phase diagrams of SECNO shown in Figs.~\ref{phasediagram}A and B, enabling us to estimate key parameters such as the exchange field \( H_J \), Maki parameter \( \alpha \), orbital critical field \( H_{\rm c}^{\rm orb} \), and Pauli limiting field \( H_{\rm P} = \tfrac{\sqrt{2}}{\alpha} \, H_{\rm c}^{\rm orb} \) (see Methods). For sample S1, we obtain a large exchange field \( H_J \approx -59~\mathrm{T} \), together with \( \alpha \approx 3 \), \( H_{\rm c}^{\rm orb} \approx 43~\mathrm{T} \), and \( H_{\rm P} \approx 20~\mathrm{T} \) (see Table II and Methods). For sample S2, a larger exchange field of $H_J\approx-71$~T is estimated (see Methods). 

The above analysis suggests the nickelates provide a second system---after the Chevrel-phase compounds---that can be described using the WHH model modified to include the effect of an internal exchange field~\cite{fischer1975,rossel1985}. The angular dependence of the phase diagram (Figs.~\ref{phasediagram}C and D) further supports this description: its $\theta$-dependence is well captured by the Tinkham model~\cite{tinkham1996} (see Methods) for the anisotropy of the orbital critical field in a two-dimensional superconductor, with both \(H_J\) and \(H_{\rm P}\) held fixed. Using the Tinkham model for \( H_{\rm c}^{\rm orb} \)---appropriate for systems with a very short interlayer coherence length~\cite{talantsev2023}---the experimental phase boundaries delineated as the midpoints of the resistive transitions are reproduced by introducing only one additional parameter: the in-plane orbital critical field, \(H_{{\rm c},\parallel}^{\rm orb} \approx 72~\mathrm{T}\) and \(\approx 69~\mathrm{T}\), respectively, in samples S1 and S2. Note that for $H\perp c$, the observation that the region of vanishing resistivity is centered near \( H \approx 30~\mathrm{T} \) provides a lower bound on \(|H_J|\) that is independent of the WHH modeling.

\begin{figure*}[!htt] 
\begin{center}
\includegraphics[width=1.0\linewidth]{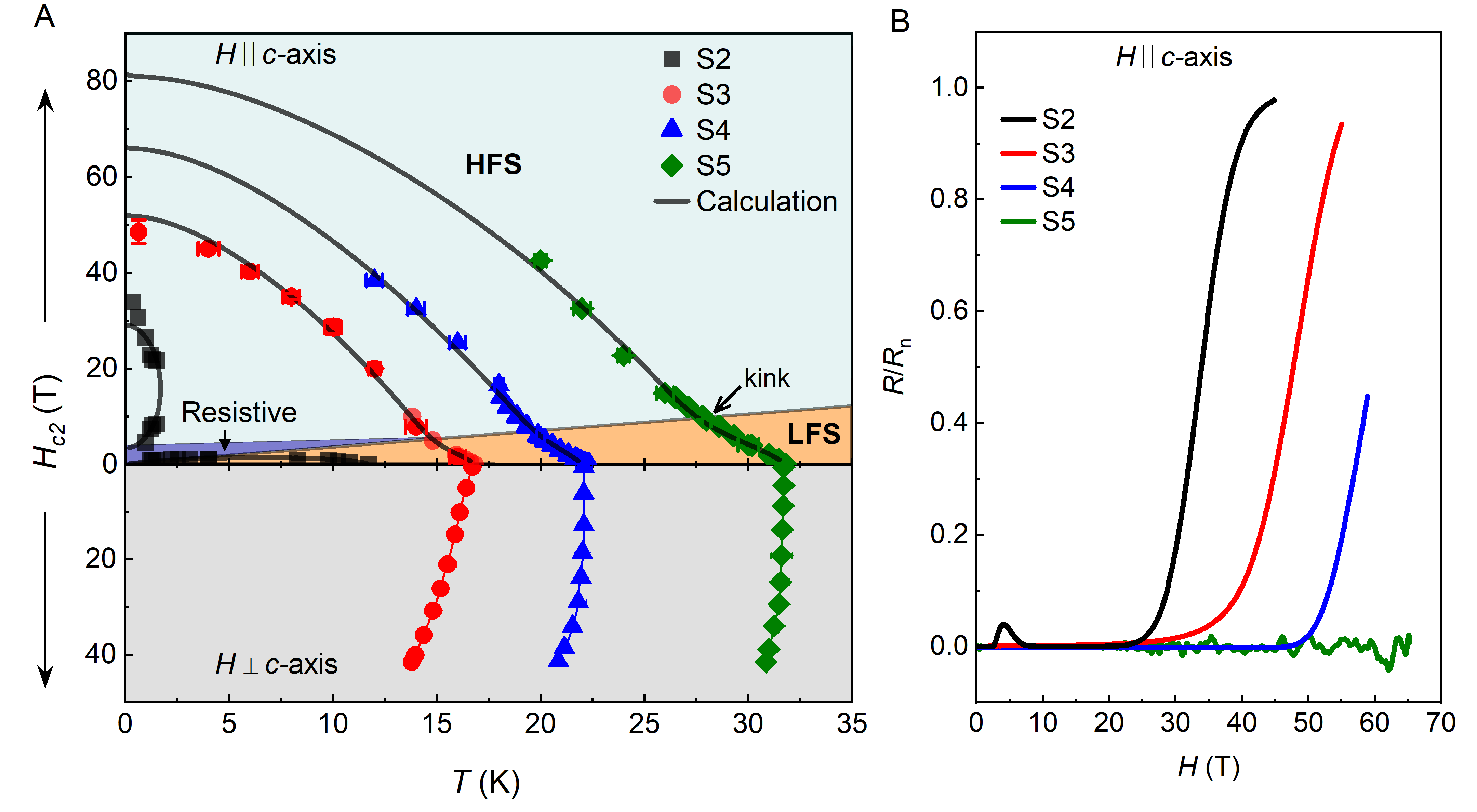}
\textsf{\caption{
{\textbf{Comparison of upper critical fields of several SECNO samples with different $T_{\rm c}$ for \(H\) $\parallel$ \(c\) axis and \(H\) $\perp$ \(c\) axis.} (A) Upper panel: Comparison of experimentally measured critical fields (symbols) for samples S2 ($T_{\rm c}=$~11.7~K), S3 ($T_{\rm c}=$~16.8~K), S4 ($T_{\rm c}=$~22.1~K), S5 ($T_{\rm c}=$~31.7~K) with phase boundaries (lines) calculated using a modified WHH model for \(H\) $\parallel$ \(c\) axis (see main text and Methods for details). Fitting parameters are listed in Table II in the Methods. Lower panel: Experimentally measured critical fields of S3, S4, and S5 for \(H\) $\perp$ \(c\) axis. In the upper panel, the orange background denotes the low-field superconducting (LFS) state, while the light-teal background represents the high-field-induced superconducting (HFS) state. The royal-blue region marks the resistive state that appears at intermediate fields for lower-$T_{\rm c}$ samples. The boundary line between the LFS and HFS regions highlights the emergence of a kink in the phase diagram. (B) Resistance renormalized by the normal state value ($R/R_{\rm n}$) measured at low temperatures of 0.35~K for S2, 0.6 ~K for S3, 0.6~K for S4 and 1.5~K for S5 for \(H\) $\parallel$ \(c\) axis. 
}}
\label{higherdopings}}
\end{center}
\vspace{-0.2cm}
\end{figure*}

The low- and high-magnetic-field–induced superconducting states merge not only when $H$ is rotated into the planes but also when the transition temperature is increased by changing rare-earth substitution or varying film thickness, as demonstrated by the evolution of the phase diagram with increasing $T_{\rm c}$ (from 11.7~K to 31.7~K) in the upper panel of Fig.~\ref{higherdopings}A. The exact compositions and film thicknesses of all studied samples are listed in Table~\ref{table1}. The raw data from which the $T_{\rm c}$ and $H_{\rm c}$ points are extracted are shown in Extended Data Figs.~\ref{RofT2} and~\ref{RofT}. In the higher-\(T_{\rm c}\) samples, this merging manifests as a kink in the phase boundary at fields below 10~T in the magnetic field–temperature (\(H\)–\(T\)) phase diagrams (upper panel of Fig.~\ref{higherdopings}A)—a feature also observed in Chevrel-phase compounds with higher transition temperatures~\cite{rossel1985}. Because this kink is fully reproduced by the WHH model (curved lines in Fig.~\ref{higherdopings}A, see Methods), its presence provides a useful indicator of the crossover between distinct low-field (LFS) and high-magnetic-field–induced superconducting (HFS) states. For fields parallel to the \(c\)-axis and \(T_{\rm c} = 31.7~\mathrm{K}\), the high-magnetic-field–induced superconducting state extends at least up to 65~T (Fig.~\ref{higherdopings}B), as evidenced by the persistence of vanishing electrical resistivity up to this field.

Upon rotating the magnetic field into the planes ($H \perp c$), where the orbital critical field is expected to be larger, we find that the internal exchange field allows superconductivity to persist to extraordinarily high magnetic fields in the higher-$T_{\rm c}$ samples. The lower panel of Fig.~\ref{higherdopings}A and Extended Data Fig.~\ref{inplane} show that increasing the magnetic field from zero to 42~T—above the Pauli limit (see Table~II)—suppresses $T_{\rm c}$ by a mere $\sim$1~K in samples S4 ($T_{\rm c}= 22.1~\mathrm{K}$) and S5 ($T_{\rm c}= 31.7~\mathrm{K}$), indicating that superconductivity is boosted to magnetic fields far exceeding the Pauli limit.

In support of a Jaccarino–Peter effect mechanism involving Eu magnetic moments in SECNO, analogous to that in Chevrel-phase compounds, our X-ray absorption spectroscopy measurements reveal that approximately 67\% of the Eu sites are in the Eu$^{2+}$ configuration (see Extended Data Fig.~\ref{eu2plus}), in agreement with Ref.~\cite{wei2023}, suggesting a static mixed-valence state. Specifically, we find that the phase transition out of the low-magnetic-field superconducting phase (see Figs.~\ref{phasediagram} and~\ref{higherdopings}, and Methods) is reproduced when using $m_J = \pm\tfrac{7}{2}$ for Eu$^{2+}$ in the WHH model. Moreover, the form of the anomalous Hall effect in sample~S1 measured above $T_c$ closely matches the magnetization calculated from the Brillouin function for $m_J = \pm\tfrac{7}{2}$ of Eu$^{2+}$ at the same temperatures (see Extended Data Figs.~\ref{magcurves}A and~B). These results suggest that the tetragonal symmetry of the lattice produces a weak crystal electric field splitting at the Eu$^{2+}$ sites. The close correspondence between the anomalous Hall effect and the Brillouin function in Extended Data Figs.~\ref{magcurves}A and~B also indicates that there is no experimental evidence for ferromagnetism involving Eu$^{2+}$ above $T_c$, although this does not preclude the possibility of a low-field magnetic state coexisting with superconductivity below $T_c$. Furthermore, while other rare-earth magnetic moments could in principle influence the phase diagram, their effect is expected to be much weaker than that of Eu$^{2+}$ (see Methods).

Given the many reported parallels between the phase diagrams of the nickelates and cuprates~\cite{botana2020,chow2022}, our finding that the nickelate phase diagram can be described by the WHH model provides important clues to the pairing mechanism. Since the WHH model, which successfully reproduces the data, is based on Bardeen–Cooper–Schrieffer (BCS) theory~\cite{bardeen1957}, this suggests that the superconducting pairing symmetry is even (i.e. non-triplet) and the BCS theory provides a useful starting point for understanding superconductivity in the nickelates. Nevertheless, open questions remain---for example, whether the theory must be revised to account for a possible unconventional $s$- or $d$-wave pairing symmetry, non-phonon-mediated pairing mechanisms, or strong fluctuation effects~\cite{Puphal2025_nickelates}.


Given the experimental evidence for high–magnetic–field–induced superconductivity in SECNO, an important question arises: why have similar effects not been observed in other high-temperature superconductors, such as the cuprates or iron–based compounds? As in the nickelates, rare-earth elements are sometimes present in their spacer layers. In the cuprates, Gd$^{3+}$ has the same magnetic configuration as Eu$^{2+}$, although magnetic-field–induced (or enhanced) superconductivity has not been reported in Gd-containing cuprates~\cite{dunlap1988}. Meanwhile, in Eu$^{2+}$-containing Fe-based superconductors~\cite{he2010,lohle2017}, only a weak magnetic-field enhancement of superconductivity has been observed, and it does not arise via a compensation mechanism. One crucial distinction between the nickelates and both the cuprates and Fe-based superconductors is the participation of the rare-earth $5d$ electrons in the conduction bands~\cite{kapeghian2020}. This provides a plausible exchange path, $4f \leftrightarrow 5d \leftrightarrow 3d_{\mathrm{Ni}}$, linking the transition-metal $3d$ conduction electrons and the rare-earth $4f$ magnetic moments~\cite{jensen1991}---an interaction that is unlikely to exist in the cuprates or Fe-based superconductors. Whether this coupling is sufficient to produce the observed magnitude of the exchange field (with $|H_J|$ as high as 70~T) remains an open theoretical question.

In summary, we have discovered high–magnetic-field–reentrant superconductivity in an infinite-layer nickelate system SECNO. Unlike previously known reentrant superconductors, whose low-field superconducting states typically exhibit low critical temperatures ($T_{\rm c}<4~\mathrm{K}$)~\cite{meul1984,uji2001,levy2005,ran2019,cao2021}, SECNO displays superconductivity with $T_{\rm c}$ values reaching up to $40~\mathrm{K}$ at ambient pressure~\cite{chow2025}, comparable to that in several high-$T_{\rm c}$ cuprate families~\cite{keimer2015}. Our measurements on multiple SECNO samples with varying $T_{\rm c}$ reveal magnetic-field–reentrant superconductivity in samples with low $T_{\rm c}$ (9.6–11.7~K), and evidence for a crossover between low-field and high-field superconducting states in samples with higher $T_{\rm c}$ (extending up to 31.7~K). In the latter, extraordinarily large critical magnetic fields are observed well beyond the Pauli paramagnetic limit (as estimated using the WHH model). In all cases, the behavior is consistent with a Jaccarino–Peter compensation mechanism. These findings outline a concrete pathway toward superconductors that operate at tens of tesla and establish high-$T_c$ nickelates with engineered rare-earth magnetism as a realistic platform for ultra-high-field superconducting magnets, sensors, and quantum devices—qualitatively different from previous low-$T_c$ reentrant superconductors.


\bibliographystyle{apsrev4-1}



\bibliography{Manuscript_v5_Dec2025/nickelateJPrevision003}



\section{Methods} 

\textbf{Sample Growth and characterization} \\
The infinite-layer nickelates Sm$_{1-x-y-z}$Eu$_{x}$Ca$_y$Sr$_z$NiO$_2$ thin films (thickness  $\approx$ 4-7 nm) were synthesized on NdGaO$_3$ (110) substrates following the method described in~\cite{chow2025} along with the improved topotactic reduction process (exact composition and film thickness of samples we studied are given in Table~\ref{table1}). In particular, we have established a setup which allows in-situ resistivity (I-V) monitoring during the reduction process. From there, we can precisely determine the reduction time for achieving optimally reduced infinite-layer phase. The schematic of this setup and corresponding resistivity with time for sample S3 is shown below in Extended Data Figs.~\ref{sampleprep}A and B, respectively. As shown in Extended Data Fig.~\ref{sampleprep}A, the as-grown sample was attached to the heater chip of the Linkam system using silver paste and dried with a low-temperature profile (60-120$^\circ$C) for approximately 35 minutes. 
Four-point contacts electrodes were made using an ultrasonic Al wire-bonder for in-situ resistance monitoring with a LabVIEW program connected to a Keithley 2400 source meter. CaH$_2$ powder was ground and placed in tandem with the mounting sample in the Linkam chamber, which was evacuated before running the reduction heating profile at 310$^\circ$C. 

During the reduction process, the crystal structure evolves gradually, while the resistance initially increases with temperature, reaching $\sim$~10$^7\Omega$, and then gradually decreases to a plateau as depicted in Extended Data Fig.~\ref{sampleprep}B. Based on our experimental experience, this plateau corresponds to the completion of the reduction process. The reduction time depends on the as-grown film quality, sample geometry, and film thickness, and may vary between samples. This means our latest reduction method is critical for optimally reduced samples such as Sm$_{0.53}$Eu$_{0.4}$Ca$_{0.07}$NiO$_2$.
Extended Data Fig.~\ref{sampleprep}C shows the XRD data of Sm$_{0.53}$Eu$_{0.4}$Ca$_{0.07}$NiO$_2$ (sample S3) during reduction optimization at different reduction stages: under-reduced and optimally reduced. The data clearly illustrates the structural evolution from the perovskite to the infinite-layer (IL) phase via a topotactic reduction. During heating with CaH$_2$, hydrogen is generated, gradually removing the apical oxygen from the perovskite lattice and hence the c-axis will shrink. In the under-reduced sample (magenta curve), the (001) peak of perovskite ($2\theta\sim$ 24$^\circ$) coexisted with the (001) peak of the infinite-layer phase ($2\theta\sim$ 27$^\circ$), indicating that apical oxygen is only partially removed. At this stage, the sample is not homogeneous. In the optimally reduced sample (purple curve), the perovskite phase is fully converted to the infinite-layer structure, with no detectable remnants from perovskite phase, demonstrating the completion of the reduction process. \textcolor{red}{The film thicknesses are estimated from the number of deposition laser pulses, with the growth rate calibrated by cross-sectional scanning transmission electron microscopy (STEM) in terms of deposition pulses per unit thickness \cite{zeng2024}}.

The x-ray absorption spectroscopy (XAS) was measured using linearly polarized x rays from the soft x-ray–ultraviolet (SUV) beamline at Singapore Synchrotron Light Source (SSLS). A total electron yield (TEY) detection method was used during the measurements. 

\begin{table*}[!htp]
\centering
\begin{tabular*}{\textwidth}{c @{\extracolsep{\fill}} c c c c}
 \hline\hline
Sample&$T_{\rm c}$(K)  &Film thickness (nm)& Composition\\
 \hline\hline
S1& 9.6&6.5(0.5)& Sm$_{0.58}$Eu$_{0.35}$Ca$_{0.07}$NiO$_2$\\
\hline
S2&11.7&4.5(0.5)& Sm$_{0.53}$Eu$_{0.4}$Ca$_{0.07}$NiO$_2$\\
 \hline
S3& 16.8&6.5(0.5)& Sm$_{0.53}$Eu$_{0.4}$Ca$_{0.07}$NiO$_2$\\
 \hline
S4&22.1&4.5(0.5)& Sm$_{0.79}$Eu$_{0.12}$Ca$_{0.04}$Sr$_{0.05}$NiO$_2$\\
 \hline
S5&31.7&4.5(0.5)&Sm$_{0.73}$Eu$_{0.2}$Ca$_{0.07}$NiO$_2$\\
 \hline\hline
\end{tabular*}
\label{table1}
\caption{\textbf {Compositional details for SECNO}. Here, $T_{\rm c}$ refers to the midpoint of the transition. Numbers in parentheses refer to the experimental uncertainty.}
\end{table*}

\textbf{Measurements} \\
Samples were wire bonded using an ultrasonic wire bonder for electrical transport measurements. High-field transport measurements were performed at both the DC and pulsed-field facilities of the National High Magnetic Field Laboratory. Pulsed-field measurements in temperature range 4 K - 0.5 K were conducted using a 60 T `Midpulse' pulsed magnet with a pulse duration of $\sim$ 500 ms and a 70 T `Duplex' pulsed magnet with a pulse duration of $\sim$ 60 ms. High DC-field measurements were  carried out using a `45 T Hybrid' magnet for temperatures below 0.5 K and using a `42 T Resistive' magnet for temperatures 1.5 K - 35 K. Additional measurements in DC magnetic fields up to 14 T were performed using a PPMS from Quantum Design Inc at temperatures above 1.5 K and in a dilution fridge at temperatures down to 0.1 K. To minimize heating induced by eddy current and magnetocaloric effect during pulsed-field measurements, we used samples of size $\sim$ 0.5 $\times$ 0.5 mm$^2$ and adopted the van der Pauw method. Measurements above 4 K were done in $^4$He gas in DC magnetic fields. Measurements between 4 K and 1.5 K were conducted by immersing the samples in liquid $^4$He, while those below 1.5 K were performed in liquid $^3$He or in liquid mixture of $^3$He and $^4$He.  A standard lock-in technique was used for all transport measurements. Home-made rotators were used for in-situ rotation of samples in magnetic fields. 

The Hall resistance (Extended Data Fig.~\ref{magcurves}A) was measured using the van der Pauw method in a PPMS by applying an excitation current of 10 $\mu$A and measuring the voltage across the diagonal contacts.

The magnetization of the 500~$\mu$m thick NdGaO$_3$ substrate along the $<$110$>$ orientation in Extended Data Fig.~\ref{magcurves}C was measured in pulsed magnetic fields using a susceptometer constructed from counterwound coils and in DC magnetic fields using a SQUID magnetometry. The DC magnetization was measured using a Quantum Design Magnetic Properties Measurement System (MPMS) with an $^3$He insert. The sample was mounted with Dow-Corning high-vacuum grease on a copper sample holder of uniform density along the axis of transport inside a plastic drinking straw lined with copper wires for thermal stability. The entire magnetically uniform $^3$He sample chamber is moved through the SQUID pickup coils by the standard MPMS transport system.

Because the nickelate thin film is deposited on a magnetic NdGaO$_3$ substrate---and its volume is $\sim$~10$^5$ times  smaller than that of the substrate---conventional measurements such as magnetic susceptibility and specific heat cannot reliably isolate the film’s signal. 
For this reason, we employed tank circuit measurements. These measurements were performed in the Midpulse magnet under the same experimental conditions used for transport measurements. Here, a proximity detector tank circuit~\cite{altarawneh2009} was transformer coupled to a 6 turn pancake coil wound from 50 gauge copper wire, to which the sample could be attached, resonating at \(\approx\!20~\mathrm{MHz}\). Screening within the sample modified the inductance on the coil, leading to a diamagnetic frequency shift. To enable digitization for $\sim$~500~ms pulse of magnetic fields, the radio frequency oscillations were mixed down to a lower frequency below $\approx$~2~MHz. For the reentrant superconducting state, the observation of a diamagnetic frequency shift in the sample required subtraction of a background measurement at the same $\theta$ and temperatures without the sample present. The form of magnetic field-dependent background observed without the sample is the result of magnetoresistance in the copper wire forming the coil that itself causes the tank circuit frequency to change with magnetic field.

\textbf{ Theoretical model} \\
We employ the Werthamer–Helfand–Hohenberg (WHH) model ~\cite{werthamer1966}, a microscopic theory that describes the upper critical magnetic field of conventional type-II superconductors in the dirty limit, incorporating the influence of localized magnetic impurities to both orbital and paramagnetic pair-breaking effects and to spin-orbit scattering. The implicit equation we used to fit experimental $H_{\rm c}$ as a function of temperature (Fig.~\ref{phasediagram}A) is \cite{werthamer1966, rossel1985}

\begin{eqnarray}\label{WHHfunction}
ln\frac{1}{t}
&=&\left(\frac{1}{2}+\frac{i(\lambda_{\rm so}-\lambda_m)}{4\lambda}\right)\times \psi\left(\frac{1}{2}+\frac{h+\lambda_m+\frac{1}{2}(\lambda_{\rm so}-\lambda_m)+i\lambda}{2t}\right) \nonumber\\
&& +\left(\frac{1}{2}-\frac{i(\lambda_{\rm so}-\lambda_m)}{4\lambda}\right) \times \psi\left(\frac{1}{2}+\frac{h+\lambda_m+\frac{1}{2}(\lambda_{\rm so}-\lambda_m)-i\lambda}{2t}\right) - \psi \left(\frac{1}{2}\right)
\end{eqnarray}
where $\lambda =  \left[ \alpha^2(h+h_J)^2-\frac{1}{4}(\lambda_{\rm so}-\lambda_m)^2\ \right]^{1/2}$.
The parameters formulated in reduced units are: $t = \frac{T}{T_{\rm c}}$, $h = 0.281 \frac{H_{\rm c}(T)}{H_{{\rm c}}^{{\rm orb}}}$, and $h_J = 0.281 \frac{H_J(T)}{H_{{\rm c}}^{{\rm orb}}}$, where $T_{\rm c}$ is the critical SC transition temperature, $H_{{\rm c}}^{{\rm orb}}$ is the orbital-limited critical field at absolute zero temperature (for $H$ perpendicular to the planes),  $H_J$ is the exchange field described by a Brillouin function for Eu$^{2+}$ ions. The above numerical factor of 0.281 is obtained theoretically by solving the gap equation in the presence of Pauli limiting (in the limit $\lambda_{\rm so}\rightarrow0$)~\cite{werthamer1966}. Meanwhile, $\lambda_{\rm so}$ and $\lambda_{m}$ are the spin-orbit and magnetic scattering parameters, respectively. $\psi$ is the digamma function with a complex argument, and $\alpha$ is the Maki parameter given by $\alpha = \sqrt{2}\frac{H_{{\rm c}}^{{\rm orb}}}{H_{P}} $ with $H_P$ being the Pauli-limited critical field. 

\begin{table*}[!htp]
\centering
\begin{tabular*}{\textwidth}{c @{\extracolsep{\fill}} c c c c c c}
 \hline\hline
Sample & $H_{\rm c}^{\rm orb}$(T)  &$\lambda_{\rm so}$ & $\alpha$ & $H_J$(T)& $H_{\rm P}$(T) \\
 \hline\hline
S1 & 43.1(0.3) & 3.7(0.4)    & 3.00(0.10) &  - 58.7(0.4)&20.3\\
\hline
S2&  44.6(0.7) & 5.5(0.9)    & 3.00(fixed) &  - 70.7(1.2)&21.0\\
 \hline
 S3&  52(2)&24(4)    & 3.7(0.2) &  - 43(5)&20.1\\
 \hline
S4& 67(4)&  28(7)   & 3.7(0.3) &  - 49(6)&25.5  \\
 \hline
S5 &83(5) &  8.8(1.2)  & 3.0(0.3) &  - 58(5) &38.9\\
 \hline\hline
\end{tabular*}
\label{theory}
\caption{\textbf{ Phase diagram fitting parameters for SECNO}. Here, $H_{\rm c}^{\rm orb}$, $\lambda_{\rm SO}$, $\alpha$, $H_J$ and $H_{\rm P}$ refer to the orbital critical field, spin-orbit scattering parameter, Maki parameter, exchange field and Pauli critical field, respectively. Note that the $H_{\rm P}$ is determined using $\alpha=\sqrt{2}H_{\rm c}^{\rm orb}/H_{\rm P}$, and is therefore not an independent fitting parameter.}
\end{table*}

The low upper critical field $\approx$~1~T is reproduced when $H_J$ saturates very rapidly at magnetic fields lower than 1~T. Such behavior is most accurately reproduced when $m_J$ = $\pm$~7/2 for the Eu$^{2+}$ magnetic moments, suggesting the presence of weak crystal electric field effects in the nickelate films. This implies that the effective $g$-factor should also scale as $\cos\theta$ for $\theta < 90^\circ$. Below we further find that a $\cos\theta$ behavior of the effective $g$-factor enables the $\theta$-dependence of the critical field of the low magnetic field superconducting phase to be understood.

Assuming $\lambda_m = 0 $ and using the fitting parameters $H_{\rm c}^{\rm orb}$, $\lambda_{\rm so}$, $\alpha$, and $H_J$ (listed in Table~II) -- we can produce the key features of the experimental data for sample S1, including low-field and high-field superconducting states, as shown in Fig.~\ref{phasediagram}A and Fig.~\ref{phasediagram}B. For sample S2, owing to the lower temperature of its reentrant phase, allowing all parameters to vary yielded $\alpha=$~5(3) and a large uncertainty in $\lambda_{\rm so}$, but not $H_J$ (its saturated high field value) or $H_{\rm c}^{\rm orb}$. We therefore repeated the fitting, holding $\alpha$ at the same value as determined for S1, where it was well-constrained.

The steep increase in $H_{\rm c}$ below 0.6 K remains unexplained using simplified WHH model and, therefore, needs elaborate theory efforts that is beyond the scope of this work. Notably, low-temperature non-saturating  $H_{\rm c}$ has also been reported in other nickelates~\cite{wang2021, wang2023} and cuprates~\cite{mackenzie1993, hsu2021} and various mechanisms, including an interplay between superconductivity and a density wave~\cite{yu2019fragile} and the existence of multiple bands~\cite{gurevich2010}, are proposed to account for this phenomenon.\\

We find that the angle-dependent phase diagram (Fig.~\ref{phasediagram}C and D) can be reproduced by using the Tinkham formula 
\begin{equation}\label{tinkham}
\Bigg|\frac{H^{\rm orb}_{\rm c}(\theta)\cos\theta}{H_{{\rm c},\perp}^{\rm orb}}\Bigg|
+\Bigg(\frac{H^{\rm orb}_{\rm c}(\theta)\sin\theta}{H_{{\rm c},\|}^{\rm orb}}\Bigg)^2 = 1
\end{equation}
for the orbital critical field of a two-dimensional superconductor, where $H_{{\rm c},\perp}^{\rm orb}$ and $H_{{\rm c},\|}^{\rm orb}$ are the orbital critical fields perpendicular and parallel to the planes, respectively. Note also that  $\alpha(\theta)=\sqrt{2} H_{\rm c}^{\rm orb}(\theta)/H_{\rm P}$, meaning that the Maki parameter has the same angular dependence as the orbital critical field if $H_{\rm P}$ is isotropic.

For sample S1, we find $H_{{\rm c},\|}^{\rm orb}=$~71.8(0.5)~T, whereas for sample S2 we find $H_{{\rm c},\|}^{\rm orb}=$~68.8(0.5)~T. This suggests that the $c$-axis coherence length is smaller than the layer spacing in infinite-layer nickelates~\cite{talantsev2023}. In the model, we also find that both $H_{\rm P}$ and the high magnetic field saturated value of $H_J$ must be isotropic (to within $\sim$~1\%) to account for the observed $\theta$-dependence of the phase boundary of the field-induced superconducting phase. 

\textbf{ Comment on the possible involvement of other rare earth moments}\\
Sm$^{3+}$ moments (\( J = \tfrac{5}{2} \), \( g_J = \tfrac{2}{7} \)) present in the infinite-layer nickelate thin films possess a small \( g \)-factor and therefore are expected to saturate only in fields of several tens of tesla, yielding a relatively small effective magnetic moment. Meanwhile, the NdGaO$_3$ substrate, where the Nd sites adopt the Nd$^{3+}$ configuration (\( J = \tfrac{9}{2} \), \( g_J = \tfrac{8}{11} \)), orders antiferromagnetically below $\approx$~1.2~K. However, because the NdGaO$_3$ magnetization saturates via a spin flop transition at \( H \approx 2~\mathrm{T} \) (see Extended Data Fig.~\ref{magcurves}C), an exchange coupling to conduction electrons near the film--substrate interface could be a factor in the sharp dip in resistivity observed at this field at low temperatures. 

\textbf{ Acknowledgements} The work at the Los Alamos National Laboratory was supported by the Department of Energy (DoE) BES project `Science of 100 tesla'. The National High Magnetic Field Laboratory is funded by the National Science Foundation through NSF/DMR-2128556, the State of Florida and DoE. The work at NUS was supported by the Ministry of Education (MOE), Singapore, under its Tier-2 Academic Research Fund (AcRF), Grants No. MOE-T2EP50123-0013 and MOE-T2EP50124-0003, the SUSTech-NUS Joint Research Program, and by the MOE Tier-3 Grant (MOE-MOET32023-0003) ‘Quantum Geometric Advantage’. Hall resistance and SQUID magnetometry measurements in moderate fields were supported by the DoE-BES through 'quantum fluctuations in narrow band systems' project. The authors acknowledge the Singapore Synchrotron Light Source for providing the facility necessary for conducting the research. The Laboratory is a National Research Infrastructure under the National Research Foundation, Singapore. Any opinions, findings, and conclusions or recommendations expressed in this material are those of the author(s) and do not reflect the views of National Research Foundation, Singapore. We thank Joe Thompson for assistance with SQUID magnetometry, and Caue Ribeiro, Arkady Shekhter, and Ross McDonald for their careful reading of the manuscript and valuable suggestions.
 
\textbf{ Author contributions} KR, KYY, AA, and NH conceived and designed the project. NF, GX, SP, and SLEC grew nickelate thin films and performed material characterization measurements under the guidance of AA. KR and EK conducted transport measurements in high pulsed magnetic fields. KR, KYY, and DG performed measurements in high dc magnetic fields. KR, KYY, and SMT measured transport in moderate fields in PPMS and dilution fridge. KR, APD, OEAV, and NH conducted magnetization measurements. KR, MKC, and NH carried out proximity-detector tank circuit measurements in high pulsed magnetic fields. CST conducted the X-ray adsorption spectroscopy measurements and GX analyzed the data under the guidance of MBHB. NH performed theoretical calculations. KR, KYY, AA, and NH analyzed the data and wrote the manuscript with input from all authors. \\

\setcounter{figure}{0}

\newpage

\begin{figure}[ht] 
 \renewcommand{\figurename}{Extended Data Fig.}
\begin{center}
\includegraphics[width=1.0\linewidth]{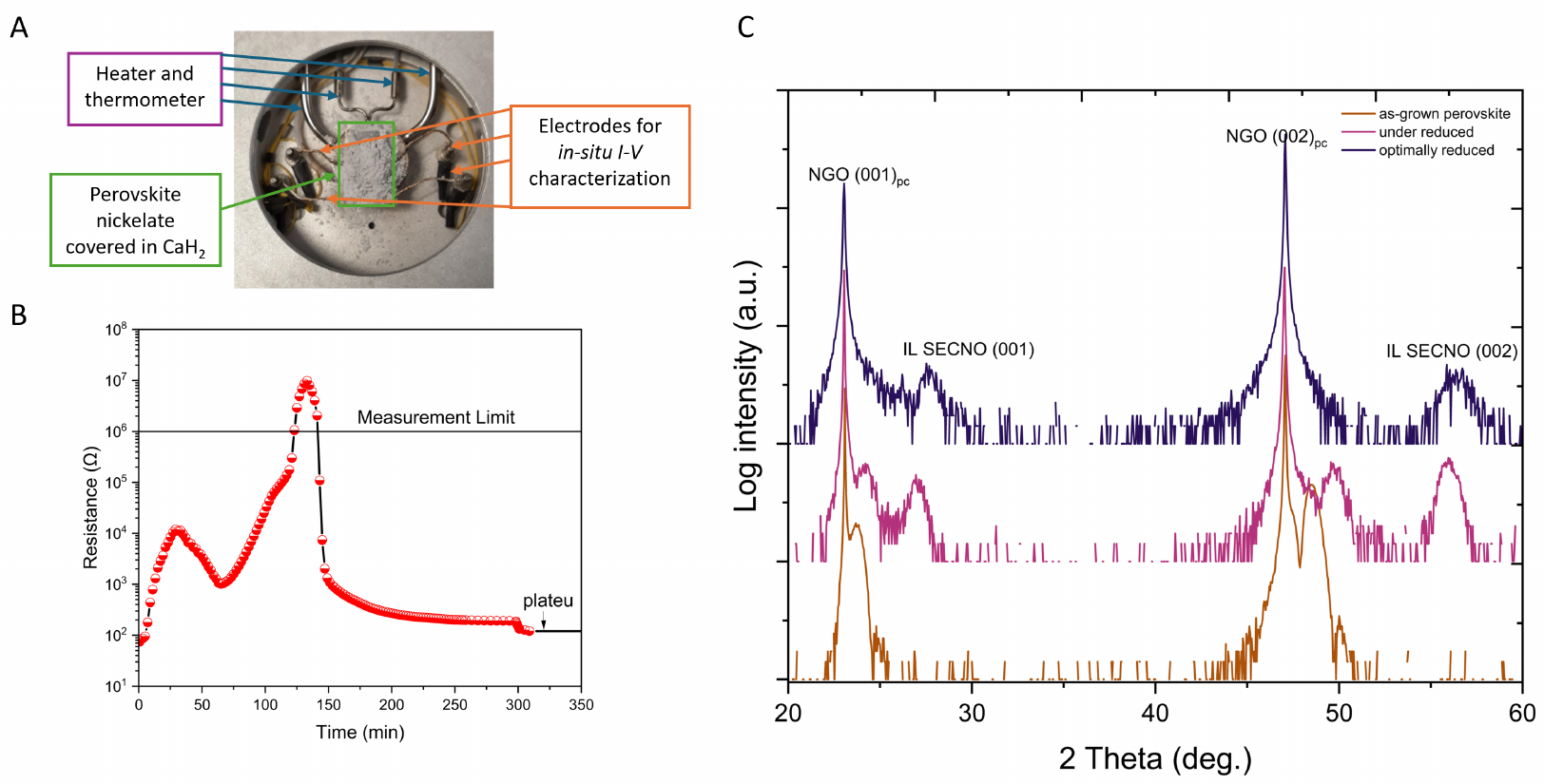}
\textsf{\caption{
{\textbf{ Sample preparation and characterization.} (A) An image of the setup for the topotactic reduction process. (B) Resistivity versus time during the reduction process. (C) XRD data of structural evolution from the perovskite to the infinite-layer phase via a topotactic reduction for sample S3.
}}
\label{sampleprep}}
\end{center}
\vspace{-0.2cm}
\end{figure}

\begin{figure}[ht] 
 \renewcommand{\figurename}{Extended Data Fig.}
\begin{center}
\includegraphics[width=0.8\linewidth]{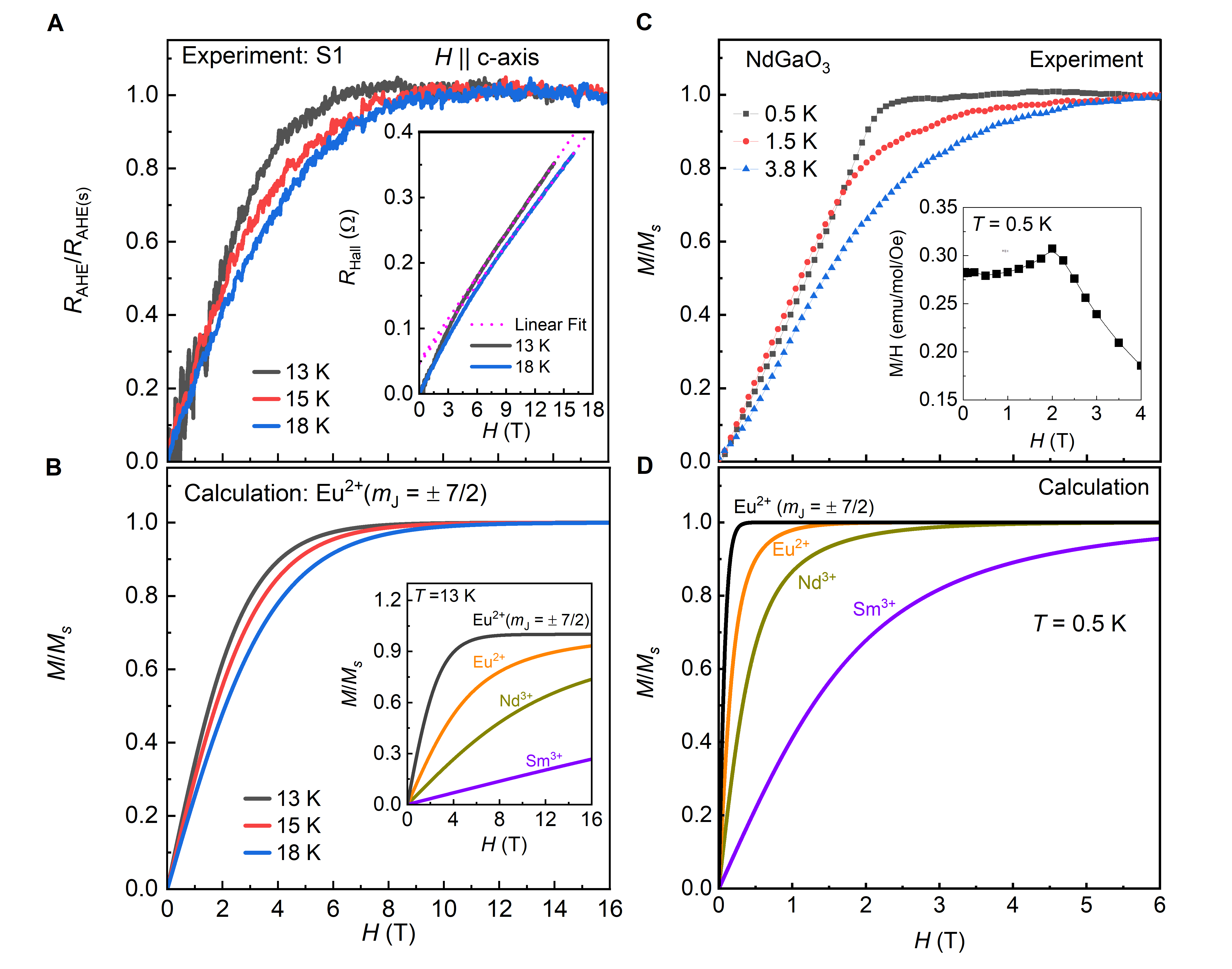}
\textsf{\caption{
{\textbf{ The anomalous Hall effect of SECNO Sample S1 and magnetization of the NdGaO$_3$ substrate with Brillouin functions.} (A) The anomalous Hall effect divided by its saturated value at three temperatures  (as indicated)  above $T_{\rm c}$ plotted versus $H$. The full Hall effect is shown in the inset. (B) The Brillouin function $M/M{\rm s}$ for $m_{\rm J}=\pm7/2$ at the same three temperatures. The inset compares this Brillouin function with those expected all of the various magnetic moments present in the thin film or substrate at $T=$~13~K, including all $m_J$ states. (C) The measured magnetization   of the substrate using pulsed magnetic fields at several different temperatures (main panel) and using SQUID magnetometry (inset), indicting a spin flop transition out of a low temperature antiferromagnetic phase at $\approx$~2~T. (D) Calculated Brillouin functions at $T=$~0.5~K. \textcolor{red}{The black curve represents the magnetization calculated by considering only the $m_{\rm J}=\pm7/2$ states, whereas the orange curve represen the total magnetization obtained by summing contributions from all possible $m_{\rm J}$ staes ( $\pm 7/2, \pm5/2, \pm 3/2, and \pm 1/2$)}.
}}
\label{magcurves}}
\end{center}
\vspace{-0.2cm}
\end{figure}	

\begin{figure}[ht] 
 \renewcommand{\figurename}{Extended Data Fig.}
 \begin{center}
 \includegraphics[width=1.0\linewidth]{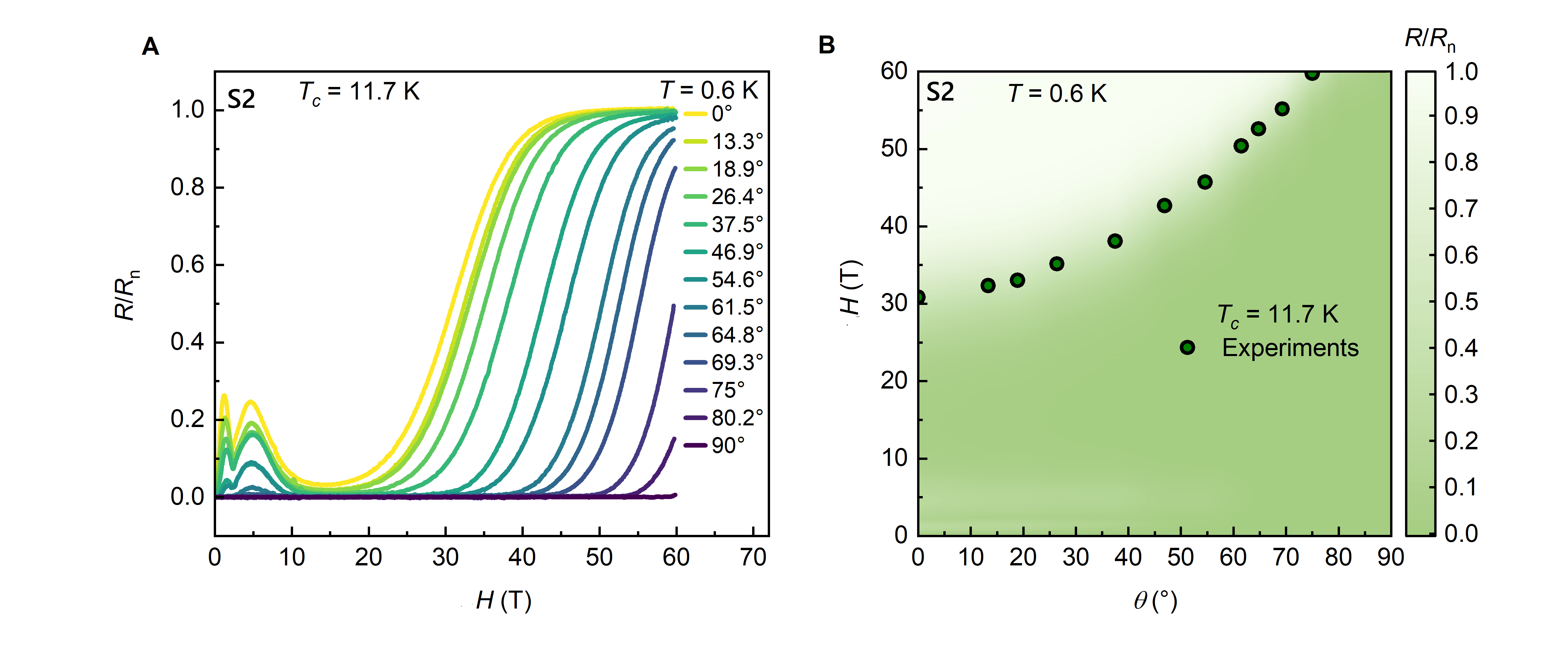}
  \textsf{\caption{
{\textbf{ Angular dependence of re-entrant superconductivity at 0.6 K for SECNO with $T_c$ = 11.7 K.} (A) Normalised resistance as a function of field at different tilt angles, and (B) Colour contour plot of the data with upper critical field values.  
}}
 \label{lowerTangular}}
\end{center}
\vspace{-0.2cm}
\end{figure}

\begin{figure}[ht] 
 \renewcommand{\figurename}{Extended Data Fig.}
\begin{center}
\includegraphics[width=0.8\linewidth]{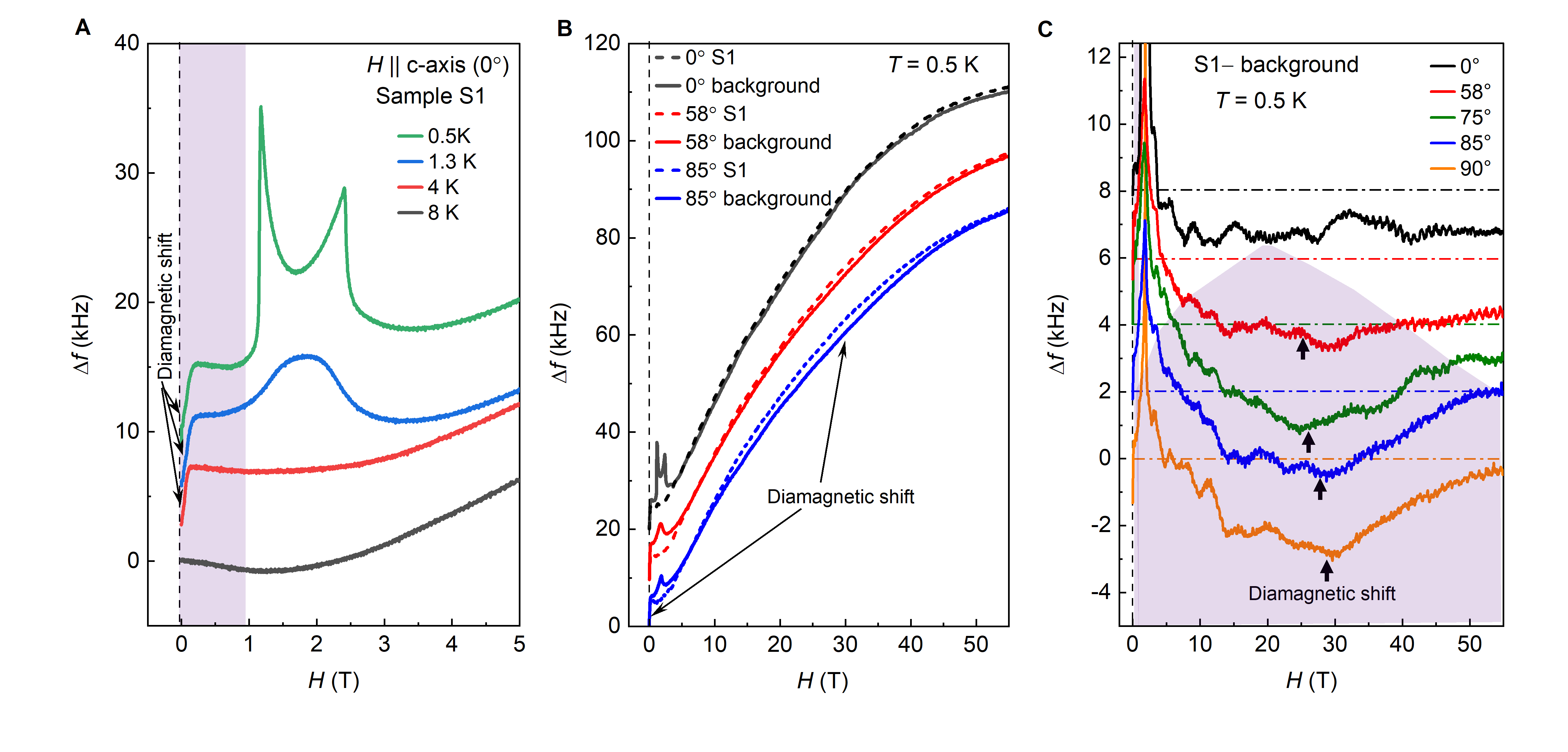}
\textsf{\caption{
{\textbf{ Tank circuit measurements of the superconducting states.} (A) Frequency shift versus magnetic field at several different temperatures in low magnetic fields. For clarity, the curves are shifted vertically with spacing of 2~kHz. Although the midpoint of resistive transition is at $H\approx$~1~T (see Fig.~\ref{resistivity}), the diamagnetic shift associated with screening currents observed as a change in frequency with field (of $\sim$~5~kHz) occurs at $\sim$~0.2~T. The broad peak at $H\approx$~2~T and $T=$~1.3~K coincides roughly with spin flop transition out of the antiferromagnetic phase of NdGaO$_3$ (see Extended Data Fig.~\ref{magcurves}). The inversion of the feature at $T=$~0.5~K coincides with the observed dip in resistivity at $H=$~2~T at temperatures 0.6~$\lesssim T\lesssim$~2~K (see Fig.~\ref{resistivity}), suggesting that some degree of interplay between the Eu$^{2+}$ moments in the nickelate film with the magnetism of the NdGaO$_3$ over this range of temperatures. (B) High magnetic field measurements with (solid) and without (dashed lines) the sample at different angles $\theta$. The curves are shifted by 10 kHz spacing for better display. Whereas the diamagnetic shift at low magnetic fields is clearly visible in the raw data, that at high magnetic fields is only discernible on subtracting a background, which is obtained by repeating the measurements without the sample \textcolor{red}{at particular temperature and angles.}. (C) Frequency shift due solely to the sample obtained by subtracting the  background at several different angles. For clarity, the curves are shifted vertically by 2 kHz, and horizontal dash-dot lines are drawn to indicate the baseline for each curve. At high fields ($\ge$ 10 T), the frequency shift for \(\theta = 0^\circ \) remains flat. In contrast, for larger angles—where the resistivity vanishes at $T=$~0.5~K (see Fig.~\ref{resistivity} and Extended Data Fig.~\ref{lowerTangular})—a clear field-induced diamagnetic shift associated with the high-field superconducting state is observed. Arrows indicate the approximate midpoints of the field range over which the resistivity vanishes. \textcolor{red}{The purple shaded regions in A and C represent the SC state observed from the transport data.} 
}}
\label{pdo}}
\end{center}
\vspace{-0.2cm}
\end{figure}

\begin{figure}[ht] 
 \renewcommand{\figurename}{Extended Data Fig.}
\begin{center}
\includegraphics[width=1.0\linewidth]{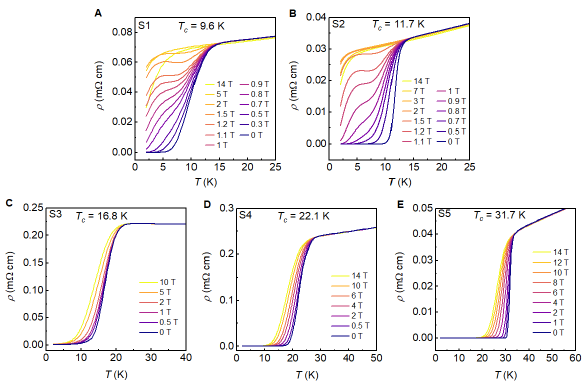}
\textsf{\caption{
{\textbf {Fixed field resistivity} Comparison of resistivity of five SECNO samples with different $T_c$, as indicated and measured at various magnetic field values. 
}}
\label{RofT2}}
\end{center}
\vspace{-0.2cm}
\end{figure}


\begin{figure}[ht] 
 \renewcommand{\figurename}{Extended Data Fig.}
\begin{center}
\includegraphics[width=1.0\linewidth]{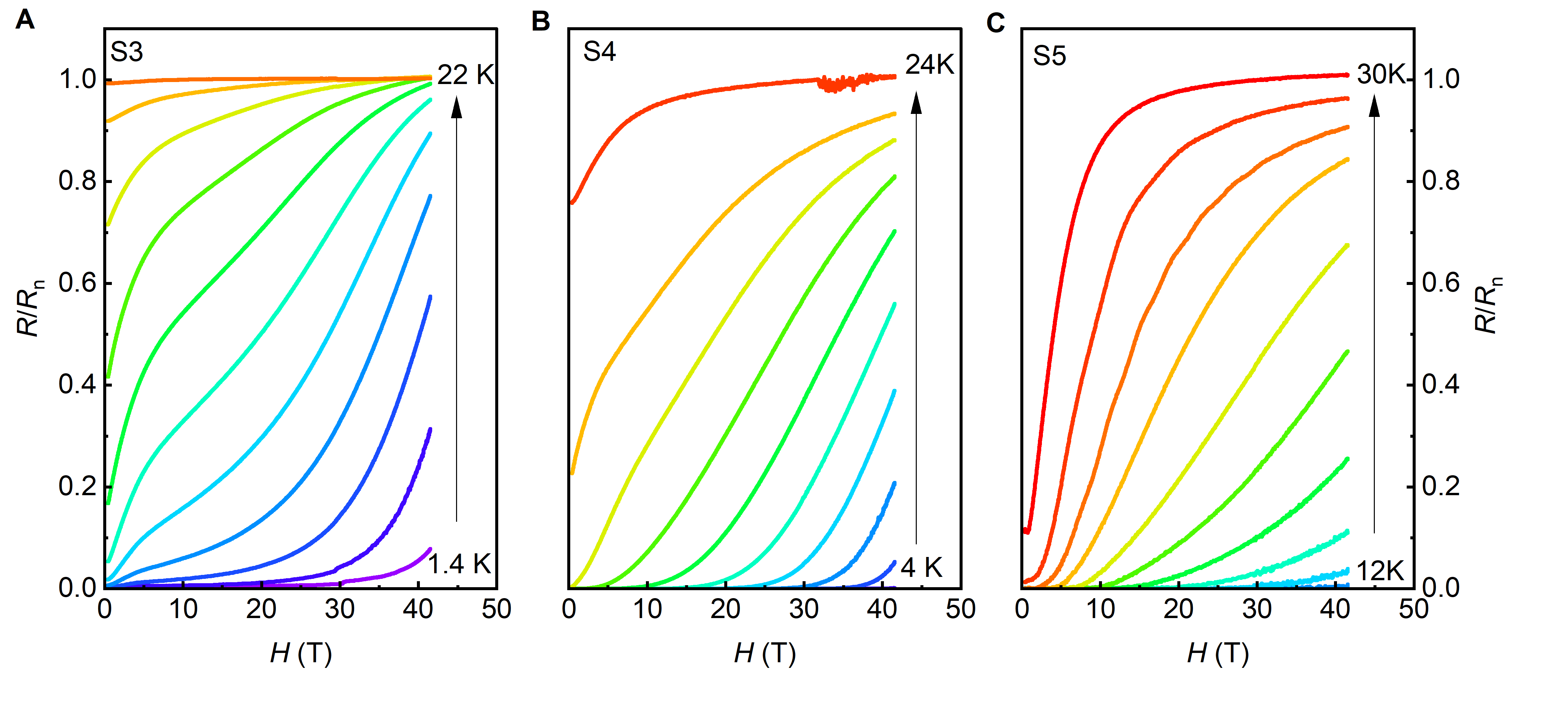}
\textsf{\caption{
{\textbf{ High-field robust superconductivity in SECNO samples S3, S4, and S5.} Resistivity sweeps of (A) S3 ($T_{\rm c}=$~16.8~K) (B) S4 ($T_{\rm c}=$~22.1~K), and (C) S5 ($T_{\rm c}=$~31.7~K) used to construct Fig.~\ref{higherdopings}A. The corresponding temperatures of the resistivity curves are 1.4~K, 4~K, 6~K, and from thereon at 2~K increments.
}}
\label{RofT}}
\end{center}
\vspace{-0.2cm}
\end{figure}


\begin{figure}[ht] 
 \renewcommand{\figurename}{Extended Data Fig.}
\begin{center}
\includegraphics[width=0.6\linewidth]{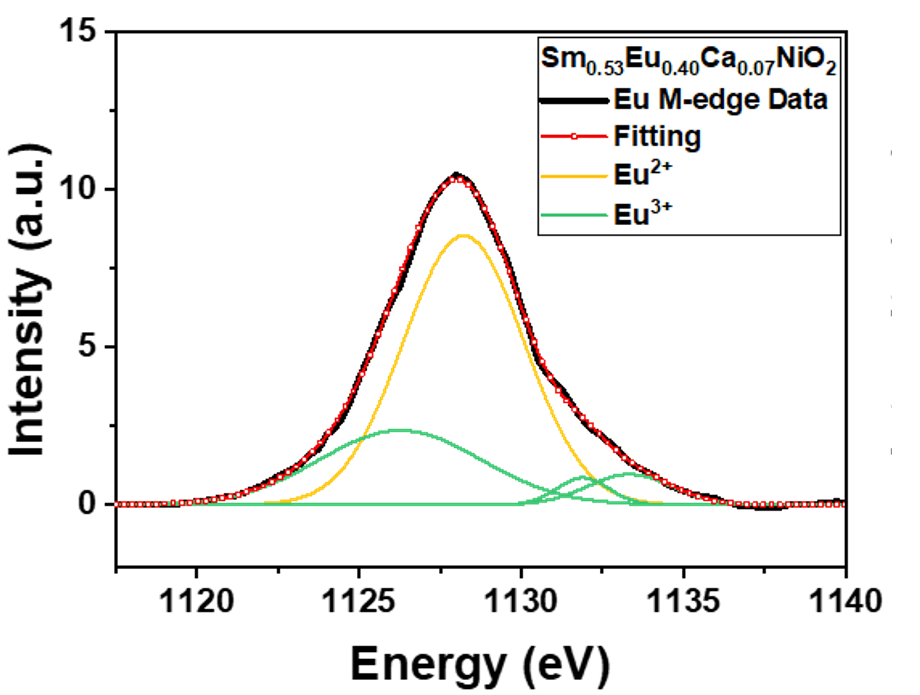}
\textsf{\caption{
{\textbf{ Evidence for Eu$^{2+}$.} X-ray absorption spectrum of Eu M-edge for S3 (see Table~1). The gaussian fitting peak positions (1128 eV for Eu$^{2+}$ and 1126, 1132, and 1134 eV for Eu$^{3+}$)  and relative areas indicate reduction of 67\% of Eu$^{3+}$ into Eu$^{2+}$ after the topotactic reduction process of the film.
}}
\label{eu2plus}}
\end{center}
\vspace{-0.2cm}
\end{figure}	

\begin{figure}[ht] 
 \renewcommand{\figurename}{Extended Data Fig.}
\begin{center}
\includegraphics[width=1.0\linewidth]{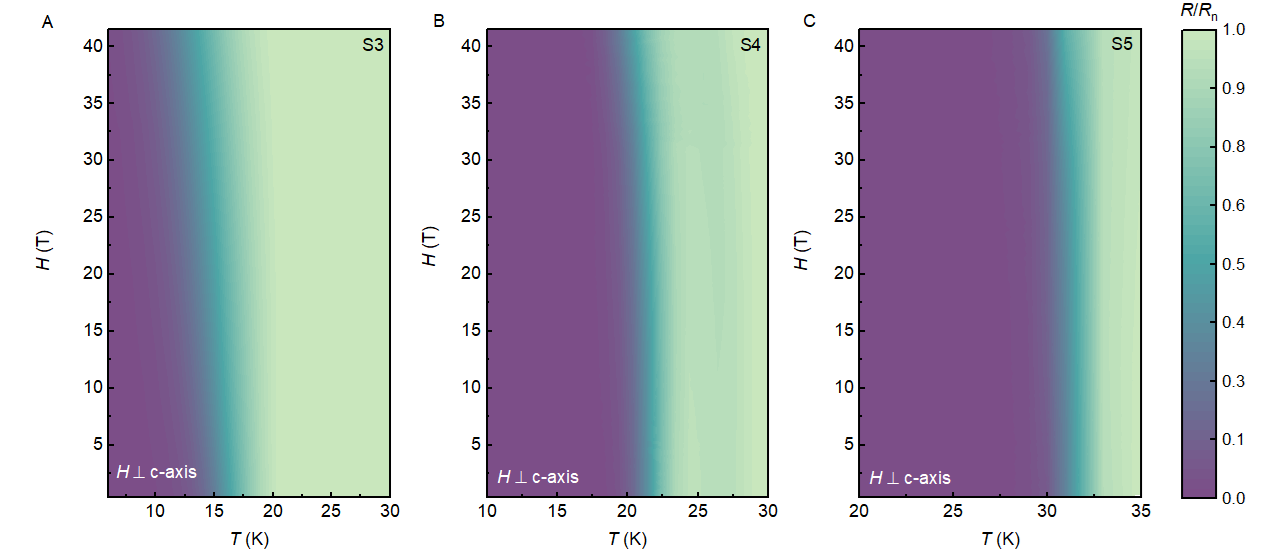}
\textsf{\caption{
{\textbf{  Phase transition measurements of high $T_c$ samples at $H\perp c$.} Colored contour plots of renormalized resistivity of (A) S3, (B) S4, and (C) S5. The data was measured while sweeping the field from 0.4 T to 42 T at fixed temperatures spaced at intervals of 1~K.
}}
\label{inplane}}
\end{center}
\vspace{-0.2cm}
\end{figure}

\end{document}